\documentclass[prl,twocolumn,superscriptaddress,showpacs]{revtex4}
\usepackage{amssymb,amsmath,graphicx,verbatim}
\newcommand{\sv}{{\scriptscriptstyle SP}}
\newcommand{\sh}{{\scriptscriptstyle SH}}
\newcommand{\re}{{\scriptscriptstyle RE}}
\newcommand{\si}{{\scriptscriptstyle SI}}
\newcommand{\sep}{{\scriptscriptstyle S}}
\newcommand{\gaus}{{\scriptscriptstyle G}}
\newcommand{\crit}{{\scriptscriptstyle K}}
\newcommand{\smm}{{\scriptscriptstyle M}}
\newcommand{\st}{{\scriptscriptstyle T}}

\newcommand{\ket}[1]{|#1\rangle}
\newcommand{\bra}[1]{\langle#1|}

\begin{document}
\title{The role of initial entanglement and nonGaussianity in the
decoherence of photon-number entangled states evolving in a noisy channel}
\author{Michele Allegra}
\affiliation{Dipartimento di Fisica Teorica, Universit\`a degli Studi
di Torino, I-10125 Torino, Italia.}
\author{Paolo Giorda}
\affiliation{ISI Foundation, I-10133 Torino, Italia.}
\author{Matteo G. A. Paris}
\affiliation{Dipartimento di Fisica, Universit\`a degli Studi di Milano,
I-20133 Milano, Italia.}
\begin{abstract}
We address the degradation of continuous variable (CV) entanglement in a
noisy channel focusing on the set of photon-number entangled states.  We
exploit several separability criteria and compare the resulting
separation times with the value of non-Gaussianity at any time, thus
showing that in the low-temperature regime: i) non-Gaussianity is a
bound for the relative entropy of entanglement and ii) Simon' criterion
provides a reliable estimate of the separation time also for nonGaussian
states. We provide several evidences supporting the conjecture that
Gaussian entanglement is the most robust against noise, i.e. it survives
longer than nonGaussian one, and that this may be a general feature for CV
systems in Markovian channels.
\end{abstract}
\maketitle
Continuous variable (CV) quantum information has been developed with Gaussian
states and operations \cite{rev1,rev2,rev3}. However, in the
recent years also the non-Gaussian sector of the Hilbert space has been
taken into consideration. This interest is due to the potential role of
non-Gaussianity in enhancing long-distance quantum communication based on
entanglement distillation \cite{ngD,ngDE} and swapping, quantum memories
\cite{ngM}, cloning \cite{ngC} and teleportation \cite{ngT}.
In turn, it has become of interest to analyze non-Gaussian states in
realistic conditions \cite{DEL}, where decoherence due to dissipation and thermal
noise unavoidably leads to degradation of entanglement.  Our work is
indeed motivated by the following general question: in case of
transmission through a noisy environment is there any advantage in using
non-Gaussian states? Do they lose entanglement in a longer time? We
provide evidence for the answer to be negative, thus supporting the
conjecture that Gaussian entanglement is extremal in terms of robustness
against decoherence due to noise and dissipation.
\par
In order to address the above questions, in this Letter we consider a
broad and meaningful class of CV bipartite states endowed with perfect
correlations in the number of photons: photon-number entangled states
(PNES). The latter have Schmidt decomposition in the Fock basis, i.e.
\begin{equation}
\ket{\psi}=\sum_{n=0}^{\infty} \psi_n \ket{n}\ket{n}
\end{equation}
with real coefficients $\psi_n \in \mathbb{R}$, $\psi_n > 0$, $ \sum_{n=0}^\infty
\psi_n^2=1$.  The advantages of considering these states are twofold.
They are sufficiently simple for analytical study, and at the same time
meaningful since several experimental realizations have been
reported~\cite{AYT}  and quantum communication schemes involving PNES
have been proposed~\cite{FUN}. Furthermore, the set of PNES contains
mostly non-Gaussian states but includes (as a subclass) two-mode
squeezed vacua, i.e.  the basic Gaussian resource for CV quantum
information, thus allowing for a direct comparison between Gaussian and
non-Gaussian states. Finally, PNES are good candidates for long-distance
quantum communication, because they have been already proved robust
against some kind of noise, e.g. phase diffusion \cite{tmw}.
We consider several special subclasses of PNES with specific parametric
dependence, as well as randomly-generated \cite{zyc98} (truncated) PNES,
in order to draw some general conclusions about the typical behaviour of
entanglement dynamics. In particular, we focus on
random PNES with decreasing profile (i.e., $\psi_n > \psi_{n+1}$) and on the
following parametric subclasses (we omit normalization): (i)  the
two-mode squeezed vacua or twin-beam states (TWB) $\psi_n \propto x^n$
$0 \leq x < 1$ which are the sole Gaussian states within the PNES class
and represent the preferred (Gaussian) resources in protocols involving
CV entanglement; (ii) the photon subtracted
(PSSV)~\cite{DAK} $\psi_n  \propto (n+1) x^{n+1}$ and the photon-added
two-mode squeezed vacua (PASV)~\cite{ZHA}
$\psi_n  \propto n x^{n-1}$, which are obtained from the TWB by the
experimentally feasible operations of photon
subtraction $\varrho \rightarrow a_1 a_2\varrho a_1^\dag a_2^\dag$ and
addition $\varrho\rightarrow a_1^\dag a_2^\dag\varrho a_1 a_2 $
respectively \cite{LNS2};
(iii) the pair-coherent or two-mode coherently correlated
states  (TMC) ~\cite{AGA} with Poissonian profile
$\psi_n \propto \frac{\lambda^n}{n!}$, $\lambda\in{\mathbb R}$.
The mean energy of PNES  is $\bra{\psi} a_1^\dag a_1 + a_2^
\dag a_2 \ket{\psi} = 2 N$ where $N=\sum_{n=0}^{\infty}
\lvert \psi_n \lvert^2 n$, whereas correlations between the modes can be
quantified by $C=\mbox{Re}\sum_{n=0}^{\infty} \psi_n^{\ast}
\psi_{n+1} (n+1)$ and entanglement is given by the Von-Neumann entropy
of the partial traces $\epsilon_0=-\sum_n \psi_n^2 \log \psi_n^2$.
In turn, the covariance matrix (CM) of a PNES equals that
of a symmetric Gaussian state in standard form, with diagonal elements
equal to $N+\frac12$ and off-diagonal blocks given by
$\boldsymbol{C}=\hbox{diag}(C,-C)$. \\
The propagation in noisy channels can be modelled as the interaction
of the two modes with two independent thermal baths of oscillators.
The resulting dynamics
is a Gaussian channel, governed by the two--mode Master equation (ME)
\begin{equation}
\dot \varrho  =  \sum_{j=1,2}\frac{\Gamma}{2}N_{j} \: L[a_{j}^{\dag}]\varrho
+\frac{\Gamma}{2}(N_{j}+1)\:L[a_{j}]\varrho
\end{equation}
describing losses and thermal hopping in presence of (local) non--classical
fluctuations of the environment. Dot stands for time--derivative and the Lindblad
superoperator is defined by $L[O]\varrho \equiv  2 O\varrho O^{\dag} - O^{\dag}
O\varrho -\varrho O^{\dag} O$. $\Gamma$ is a loss coefficient and $N_{j}$ are the mean
photon-numbers in the stationary state, which is a thermal state. We consider
baths at equal temperature $N_1 = N_2 = N_T$.
The above ME admits the operator solution \cite{gmd94}:
$\varrho(t) = \Lambda_t \varrho (0) = \mbox{Tr}_{34} [U_t (\varrho(0)
\otimes \nu_{34} ) U_t]$, where $\Lambda_t$ denotes the evolution map corresponding
to the noisy channel; $3, 4 $ are two additional fictitious modes in a thermal
state $\nu_{34} = \nu_3 \otimes \nu_4$;  $U_t=U_{13}(\zeta_t) \otimes
U_{24}(\zeta_t)$ and $U_{ij}(\zeta_t)= \exp(\zeta_t a_i^\dag a_j -
\zeta_t^* a_j^\dag a_i)$ is the two mode mixing operator,
with $\zeta_t=\arctan(e^{\Gamma t}-1)^{1/2}$. Using this solution, the
evolved density matrix $\varrho_t$ can be computed numerically from the
initial state $\varrho_0$ upon truncating the Hilbert space dimension.
In our study we consider states with total energy $ 0 \leq 2N \leq 10 $ and
dimension $D=20$. In this range of energies, and for all
subclasses of states, the adopted truncation results in a negligible error.
We emphasize that the map $\Lambda_t$, being the product of two local maps,
can only disrupt quantum correlations: for any $N_T \neq 0$ we have a complete
loss of entanglement within a finite, state dependent, time $t_{\sep} =
t_{\sep} (\varrho)$ which we refer to as the \emph{separation time}.
\par
In order to estimate $t_{\sep}$ for non-Gaussian states subjected to
the action of $\Lambda_t$ we make use of several entanglement criteria and
this also enables a comparison of their performances in detecting
entanglement. As it is well known, in the CV case a necessary-and-sufficient
separability criterion exists only for Gaussian states \cite{SIM4}:
Simon's criterion (SI) for separability is equivalent to the positivity
of the partial transpose density matrix and says that a Gaussian state is
separable iff $ \tilde{d}_{-} < 1/2  $, where $\tilde{d}_{-}$ is the least
symplectic eigenvalue of the CM of the
partial-transposed state.
When dealing with CV non-Gaussian states, Simon's criterion
(which is equivalent to the separability of a Gaussian state having the
same CM as the given state) is only sufficient for
entanglement.  This actually holds for
any available criterion: if the state is entangled,
a given test may or not detect its entanglement; in turn, if no
test detects entanglement, we can not conclude separability of the
state. The Simon separation time $t_{\si}$ can be computed
analytically. At the level of CM, the map $\Lambda_t$
induces the evolution  $ \sigma_t = \sigma_0 e^{-\Gamma t} +
\sigma_{\infty} (1 -e^{-\Gamma t})$, where $\sigma_{\infty}
=\mbox{diag} (N_\st+1/2, \dots , N_\st+1/2)$ is the asymptotic thermal
state CM. The CM of the
partial-transposed state is given by $\Lambda \sigma_t \Lambda$, where
$\Lambda =
\mbox{diag}(1,1,1,-1)$ ~\cite{SIM4}, and 
we have $\tilde{d}_{-}=
(N_\st+1/2)(1- e^{-\Gamma t}) + (N+1/2)e^{-\Gamma t}- |C|e^{-\Gamma t} $.
Therefore, for $N_\st=0$ PNES are entangled at any
time, whereas for $N_\st \neq0$ we have a lower bound to separability
\begin{equation}
t_{\si}  = \frac{1}{\Gamma} \log \left( 1 + \frac{|C|-N}{N_\st}\right).
\end{equation}
\par
Besides Simon's criterion, we will make use of three different
criteria which provide independent separability conditions.
The first is the extension of SI given by Shchukin and
Vogel~\cite{VOG1}(SH) based on the evaluation of a series of $M \times M$
matrices whose
entries are moments up to a given order: non-positive-definiteness of any
finite submatrix is a sufficient condition for entanglement.
By considering the minor defined by the first and second-order moments only ($M=5$)
we obtain a condition which is equivalent to SI. If we consider
larger minors, moments of higher order are involved and we get a stronger
condition. Here we consider moments up to order $8$.
The second criterion has been introduced by Sperling and
Vogel (SP)~\cite{VOG} and it is based on linear entanglement witnesses.
A state $\varrho$ is entangled if $\bra{\phi} \varrho \ket{\phi} >
\max_n \{ | m_n|^2 \}$ where
$|\phi\rangle$ is a pure entangled state with Schmidt coefficients $\{ m_n \}$.
We test this condition by using $10^4$ randomly generated witnesses
of the form $\ket{\phi} = \sum_{n=1}^D \phi_n \ket{n}\ket{n} $ with $D=20$, i.e.
the witnesses are themselves truncated random PNES.
This form is chosen since the bath does not create quantum correlations but
only destroys those originally present. Finally, the realignment
criterion~\cite{CHEN} (RE) is based on positivity of a linear contraction
map: a state $\varrho$ is entangled if
$|| \tilde{\varrho} || > 1$ where $ || A ||$ denotes the trace norm of operator $A$
and $\bra{m}\bra{\mu}\tilde{\varrho} \ket{n}\ket{\nu} = \bra{m}\bra{n}
\varrho \ket{\nu}\ket{\mu}$.
Using these criteria, we obtain lower bounds on separation times.
Indeed, for any given criterion $K$ and state $\varrho$, let us denote
by $t_{\crit} (\varrho)$ the maximum time for which $K$ proves that $\varrho$
is entangled: clearly $t_{\crit} (\varrho)$ is a lower bound for $t_{\sep}$.
Considering the best bound we have 
$ t_{\sep} (\varrho) \geq
t_\smm=\mbox{max}_{\crit}  t_{\crit} (\varrho). $
\par
The propagation in noisy channels, besides entanglement, also destroys
the non-Gaussian character of the initial state, which unavoidably
evolves towards the asymptotic, Gaussian thermal state. We shall take into
account both processes (separation and Gaussification) in parallel and explore
the relations between them. The non-Gaussian character
of a state  $\varrho$  is measured by 
$
\delta (\varrho)= S(\tau)- S(\varrho)
$
i.e, the relative entropy between $\varrho$ and the reference Gaussian state
$\tau$ having its same covariance matrix \cite{GPB}.
In order to explore the effect of noise in a wide range of conditions
and initial states we consider  TWB, PSSV, PASV, TMC and random PNES
of different energies and compute the evolved density matrix for
$ 0 \leq t \leq 15 $ in units of inverse loss $1/\Gamma$. At any time
$t$, entanglement is tested with all the above mentioned criteria and
the value of the non-Gaussianity $\delta$ is computed. From these data
we evaluate $t_{\crit} $, i.e. lower bounds to separation times according to
different criteria,  and Gaussification times $t_{\gaus}$, i.e. times
for which non-Gaussianity $\delta$ falls below a fixed Gaussification
threshold $\delta_{\gaus} $ (we consider
different thresholds $\delta_{\gaus} = 0.1, 0.01, 0.001$). The procedure
is then repeated for different values of the temperature $T$ corresponding to
$N_\st $ in the range $[10^{-5} , 10^{-1}]$.
\par
\begin{figure}[h!]
\includegraphics[width=0.49\columnwidth]{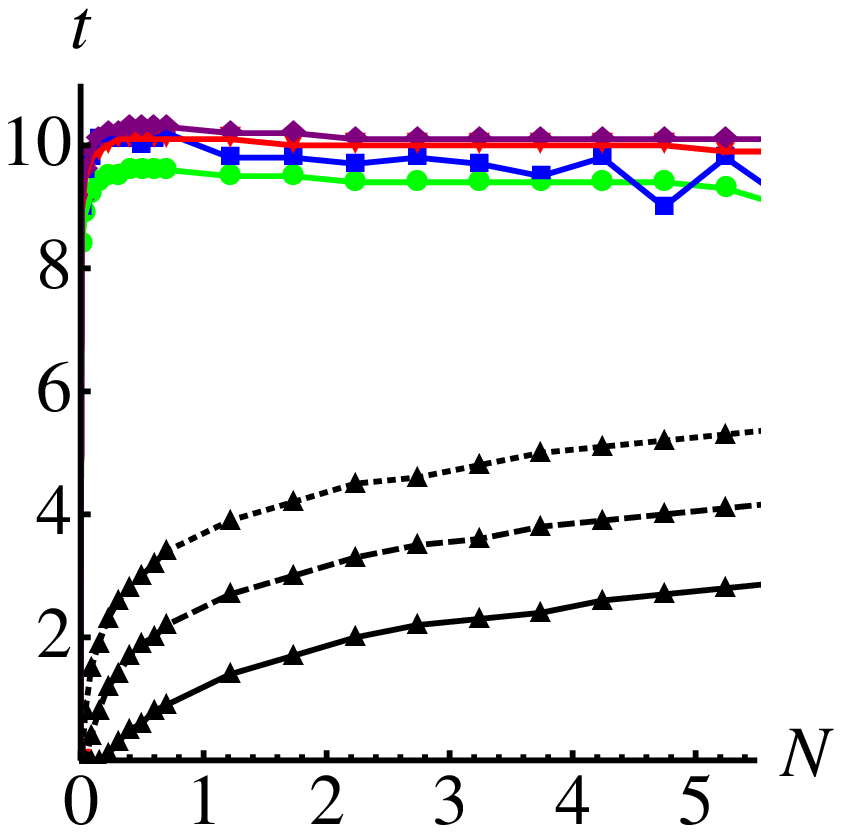}
\includegraphics[width=0.48\columnwidth]{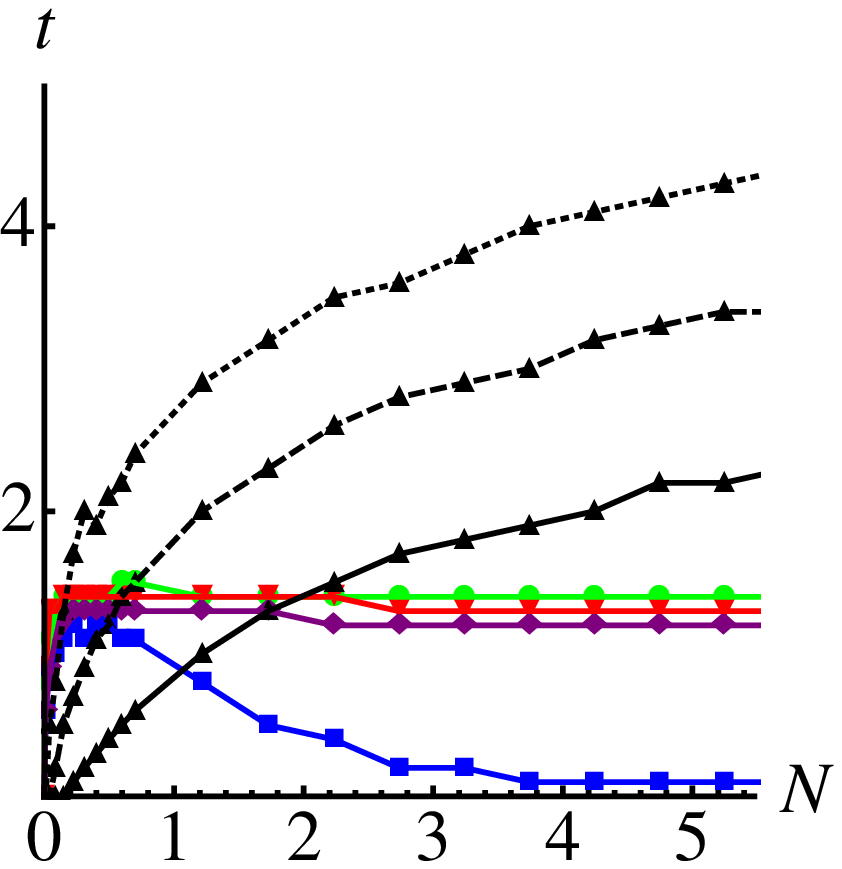}
\caption{ (Color online) Separation and Gaussification times for TMC as
a function of the mean energy for low ($N_\st =10^{-5}$, left) and high
($N_\st =10^{-1}$, right) temperature. In both plots we report
separation times according to different criteria: $t_{\re}$ (green, circle),
$t_{\sv}$ (blue, square), $t_{\sh}$ (red,triangle), $t_{\si}$ (purple,rhombus), and
Gaussification times for different thresholds: $\delta_{\gaus} =
10^{-1}$ (solid black,triangle), $\delta_{\gaus} = 10^{-2}$ (dashed black,triangle),
$\delta_{\gaus} = 10^{-3}$ (dotted black,triangle).}\label{twbsepa}
\end{figure}
We start describing the results of our analysis by focusing on TMC.
In Fig. \ref{twbsepa} we report $t_{\crit}$ for TMC and different criteria
as a function of
$N$ for the lowest (highest) temperature
considered $N_\st=10^{-5}(10^{-1})$. It turns out that at any temperature
SI, SH and RE  criteria yield similar curves whereas the SP
criterion works only at low $T$. We point out two general features:
(i) $\bar{t}_{\crit}$ is a decreasing function of $T$, i.e.
entanglement is strongly corrupted as the temperature increases;
(ii) both at high and low $T$,
$t_{\crit}$ rapidly increases to an asymptotic value $\bar{t}_{\crit}$
which is reached at $N \sim 1/2$ and then remains almost constant.
In Fig. \ref{twbsepa} we also show Gaussification times of TMC as a function
of energy. We see that the behaviour of non-Gaussianity is only weakly affected
by the increase of $T$. Upon comparing separation and Gaussification
times we notice that at low $T$ states become nearly
Gaussian well before they become separable:
$t_{\gaus} < t_{\crit} \leq t_{\sep}$.
At high $T$, on the contrary, Gaussification times are greater than our
bounds on separation times: $t_{\gaus} > t_{\crit}$. The analysis of
all other PNES subclasses (TWB, PASV, PSSV and random PNES)
reveals the same qualitative behaviour described for TMC
for the evolution of both entanglement and
non-Gaussianity (with the obvious exception of TWB, whose
non-Gaussianity is always zero).  In summary, we have numerically
proved that {\em for the whole class of states we have considered
and at any temperature SI, SH and RE
criteria yield qualitatively the same results}.
In addition, Simon's criterion, which offers analytical advantages,
is the optimal one.
Furthermore, the separation time decreases with $T$, and the
dependence on the energy can be appreciated only for small $N$, while
they quickly reach their asymptotic values as $N$ increases.  As for the
non-Gaussianity, we have $t_{\gaus} < t_{\crit} \leq t_{\sep}$ at low $T$
and $t_{\gaus} > t_{\crit}$ at high $T$.  This deserves futher
consideration.  Indeed, the relation $t_{\gaus} < t_{\crit} \leq
t_{\sep}$ suggest that for low $T$ the bounds provided by Simon's
criterion properly estimate the actual PNES separation times i.e.,
$t_{\si} \simeq t_{\sep}$.  This can be understood by first noticing
that when $t>t_{\gaus}$ the states are nearly Gaussian and therefore
Simon's criterion is expected to be very reliable. Furthermore, at any
$t \geq t_{\si}
> t_{\gaus} $ the reference Gaussian state is obviously separable
and thus the non-Gaussianity can be compared with a measure of
entanglement: the relative entropy  ~\cite{VED} $E (\varrho) =
\mbox{min}_{\sigma \in \Omega} [ S (\varrho || \sigma )] $ that
quantifies the distance between $\varrho$  and the whole set of
separable states $\Omega$.  When $t \geq t_{\si}$ one has that 
\begin{equation}
E(\varrho)\leq \delta (\varrho)\ll 1
\end{equation}
and this confirms that in this
limit the states are very poorly entangled (if they are) and SI allows
to reliably estimate $t_{\sep}$. The fact that the other criteria
provide very close bounds on $t_{\sep}$ strengthens our
conclusion.  At high $T$ Gaussification times are greater than
all $t_\crit$ and we cannot draw the same
conclusions. However, the agreement between different criteria
is still an indication that $t_{\crit}$ may represent a good estimate
of $t_\sep$.
\par
We now focus our attention on the dependence on $t_{SI}$ on the initial
non-Gaussianity $\delta_0$. Indeed, we have $\delta_0 = 2 f(d_{-})$
where $f(x)=(x+1/2)\log{(x+1/2)}-(x-1/2)\log{(x-1/2)}$ monotonically
increases with $x$ ~\cite{rev3} and $d_{-}=\sqrt{(N+1/2)^2-|C|^2}$ is
the least symplectic eigenvalue of the covariance matrix. Upon defining
$g = f^{-1}$, $t_{\si}$ can be written as 
\begin{equation}
t_{\si} = \frac{1}{\Gamma}
\log \left(1+ \frac{ \sqrt{(N+1/2)^2- g^2(\delta_0/2)}-N }{N_T}
\right)
\end{equation}
which shows that {\it $t_{\si}$ is a decreasing function of $\delta_0$
at any fixed $N$}, and it is maximized by TWB for which $\delta_0 = 0$.
\begin{figure} [hbt]
\includegraphics[width=0.49\columnwidth]{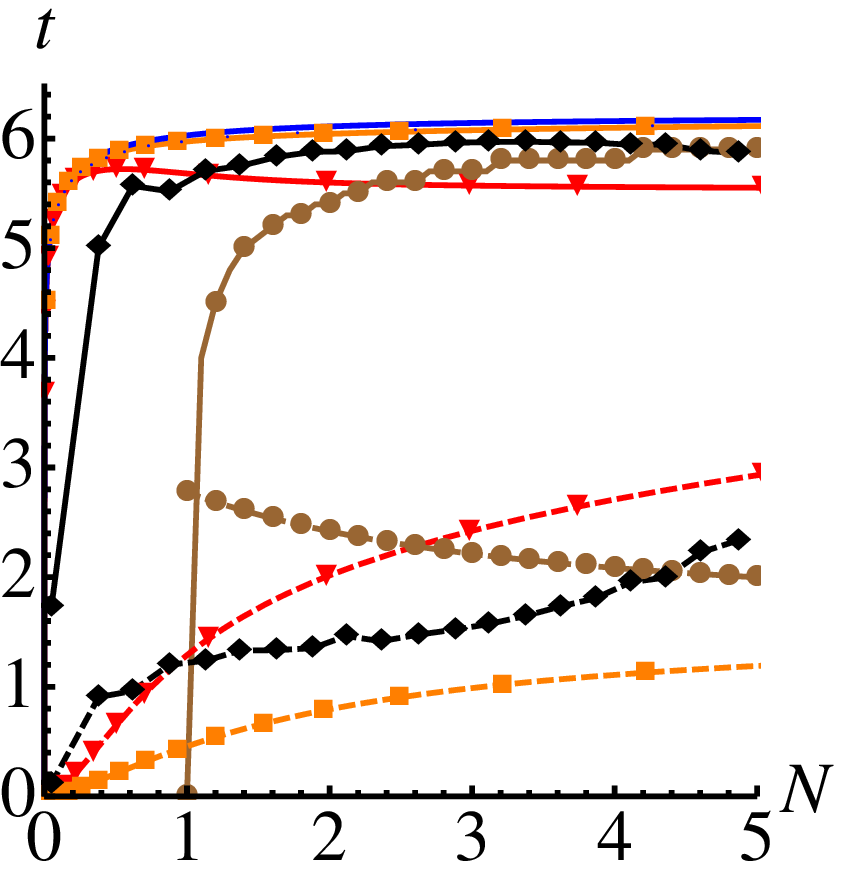}
\includegraphics[width=0.49\columnwidth]{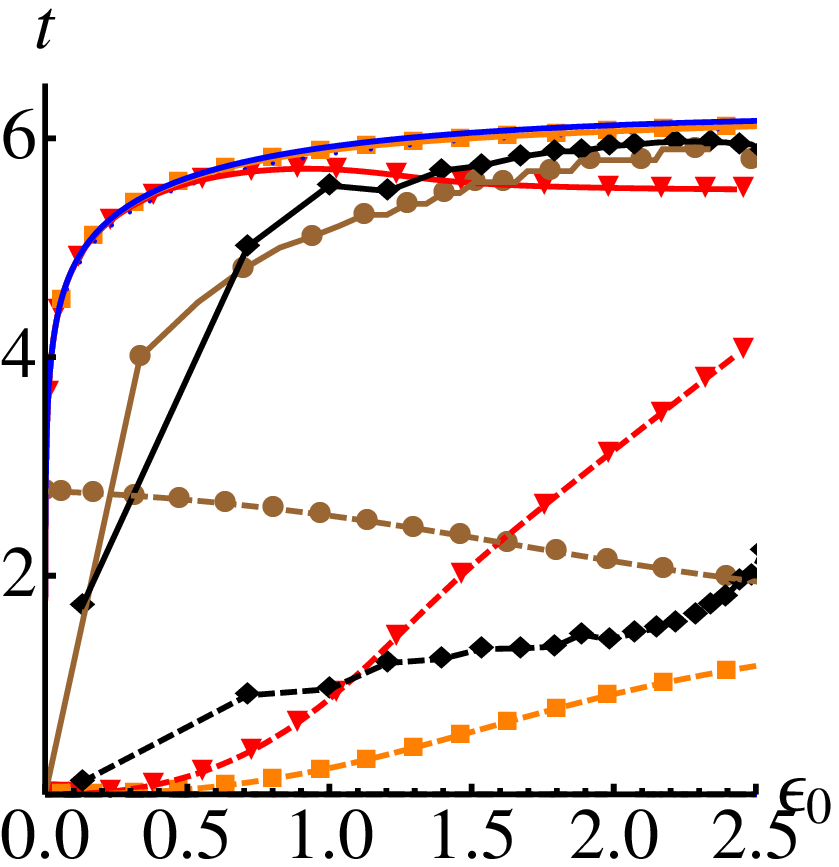}
\caption{Separation times $t_{\sep}$ (straight lines) and
initial non-Gaussianities $\delta_0$ (dashed lines) 
 as a function of initial energy $N$ (left) and initial entanglement
$\epsilon_0$ (right) for different PNES classes: PASV (brown, circle), PSSV (orange, square),
TMC (red, triangle), TWB (blue) and random (black, rhombus) states.
In both cases the bath has fixed $N_\st=10^{-3}$.} \label{sepatimes}
\end{figure}
\par
In the left panel of Fig. \ref{sepatimes} the separation times
and the initial non-Gaussianities of different PNES
are plotted against $N$ (for $t_\sep$ we use $t_\smm =\mbox{max}_{\crit}
t_{\crit} (\varrho)$): at any fixed energy $N$, the states with higher
$\delta_0$ have shorter separation times. This result holds $\forall T$
and it is related with the fact that at any fixed $N$ Gaussian states
are maximally entangled~\cite{rev3}.
Let us now discuss the relation between separation times, non-Gaussianity
and the initial entanglement of the states. As shown in the right panel
of Fig. \ref{sepatimes}, where $t_M$ and $\delta_0 $ are plotted as a
function of the initial entanglement $\epsilon_0$, the dependence is by
no means universal. However, we notice that also at fixed $\epsilon_0$
states with higher $\delta_0$ show shorter $t_\sep$: this trend is not
represented by an exact relation, but it represents a clear indication
that non-Gaussianity speeds up the loss of entanglement, making Gaussian
entanglement more robust than non-Gaussian one.
Therefore the robustness of Gaussian entanglement may be conjectured to be a general
feature of CV systems evolving in noisy Markovian
channels.
In fact, within the Markovian approximation, propagation in CV
noisy channels corresponds to a ME in Linblad form, which induces a
Gaussian map and enforces Gaussification of any initial state.
\par
The results of our analysis, together with the above discussions,
naturally lead us to formulate the following general conjecture: {\it
for any fixed value of the global energy of a PNES, and
for any given
noisy Markovian evolution with losses and thermal hopping, the Gaussian
states are those that have maximal separation times}.
Besides, from Fig.\ref{sepatimes} we also extract another relevant
feature: \emph{in the high-energy limit there is an approximate universality in
separation times}  i.e., $t_{\sep}$ are nearly constant and similar for
all classes of states, including randomly-generated states, independent
of the non-Gaussianity: the effect of the departure from Gaussianity
is very small.
\par
Let us summarize the results of our analysis. We have considered a class
of states (PNES) including Gaussian and non-Gaussian subclasses and
exploited several entanglement criteria to estimate their
separation times in a noisy channel. The analysis shows that no criterion
is able to give better bounds than those provided by Simon's criterion.
At low temperature, the estimate provided by Simon's criterion is very
reliable since PNES gaussify well before they lose entanglement,
whereas at high temperature it represents a lower bound on separation
time.  At any fixed energy $N$, separation times decrease with the initial
non-Gaussianity $\delta_0$, both at high and at low temperature,
whereas for any fixed initial entanglement $\epsilon_0$
separation time is longer for states with lower $\delta_0$, i.e. Gaussian
entanglement is the most robust against noise.
Finally, in the high energy limit and
independently of the temperature, the differences among separation times
of different subclasses are small, non-Gaussian entanglement being nearly as
robust as Gaussian one.
\par
In conclusion, we have provided several evidences supporting the
conjecture that at fixed energy Gaussian entanglement is the most robust
against noise in a Markovian Gaussian channel.  On the other hand, our
analysis shows that robustness of non-Gaussian states is comparable with
that of Gaussian states for sufficiently high energy of the states.
This implies that in these regimes non-Gaussian resources can be
exploited to improve quantum communication protocols approximately over
the same distances.
\par
MA thanks Dr. Marco Genovese for useful comments. MGAP thanks Vladyslav Usenko
for useful discussions in the early stage of this work.

\end{document}